# Development of a miniaturized motion sensor for tracking warning signs of low-back pain

Jérôme Molimard[a], Tristan Delettraz[a], Etienne Ojardias[b]

[a] Mines Saint-Etienne, Univ Lyon, Univ Jean Monnet, INSERM, U 1059 Sainbiose, Centre CIS, F - 42023 Saint-Etienne, France
[b] Clinical Gerontology Department, University Hospital of Saint-Étienne, Saint-Étienne France
[c] SAINBIOSE, INSERM U1059, University of Lyon, Saint-Étienne, France.
molimard@emse.fr, tristan.delettraz@etu.mines-stetienne.fr, Etienne.Ojardias@chu-st-etienne.fr

Abstract: Low-back pain (LBP) is a widespread disease which can also be highly disabling, but physicians lack of basic understanding and diagnosis tools. During this study, we have designed and built a new wearable device capable of detecting features helpful in LBP follow-up while being non-invasive. The device has been carefully validated, and shows good metrological features, with small noise level ($\sigma = 1°$) and no observable drift. Two simple exercises were proposed to two young volunteers, one of them with LBP history. These exercises are designed to target two characteristics: the lumbar lordosis angle and the hip & shoulder dissociation. Even if no general rules can be extracted from this study, we have shown that Inertial Measurement Units (IMU) are able to pick up those characteristics and the obtained values are meaningful refereeing to LBP disease. Henceforth, we are confident in going to clinical studies to investigate the link between back related feature and LBP, in particular the hip & shoulder dissociation which is poorly documented.

## 1. INTRODUCTION

Low-back pain (LBP) is a widespread affliction in most developed and industrialized countries. It is a major disability factor both at work and in every-day life (Bauer et al., 2017). Worldwide, it is the most reported reason for seeking care from a primary care physician (Traeger et al., 2017). LBP is accountable for the most sick leaves and it touches 70% of people at least once in his/her life (Koes et al., 2006). Therefore, it costs the French government more than one billion euros each year (Depont et al., 2010). LBP care is difficult since about 90% of all patients suffer "non-specific low-back pain" which means that while the pain is apparent, its cause remains unknown (Koes et al., 2006). Furthermore, 73% of patients experiencing LBP will go through an other episode within a year (Koes et al., 2006) which will usually be more painful than the last one (Riihimäki et al., 1991).

Hence, a lot of scientific studies have been made in order to find a way to better monitor the condition of patients afflicted by LBP, among them the lumbar lordosis angle (Evcik et al., 2003). Likewise, since people afflicted by LBP may have trouble performing normal muscle activities, like bending their back sideways. The detection and quantification of some abnormal minute back movements, like the hip & shoulder dissociation (Park et al., 2012), could establish a new method to label the beginning of a serious illness. Both subjects generated great interest in the medical community (Baek et al., 2010).

Many methods to monitor those characteristics have been put forward. For instance, Electromyography (EMG) is the most popular technique for muscle activity observation. EMGs can measure various movements at a high rate, with almost no added weight. Their most commonly reported drawbacks are the difficulties to set up the device and its sensors need to by applied sometimes invasively under the skin of the patient (Butler et al., 2010). Moreover, the results given by EMG still need to be processed to get the desired displacement or angle.

A popular alternative solution is the use of optical motion capture system. Nevertheless, the high cost of the cameras and markers may be a limiting factor to its usage, as well as the fact that it can only handle movements contained in a closed and limited area (Nakamoto et al., 2018).

Recently, strain sensors were successfully used to monitor lumbar motion. They also can be made wearable, unlike the previous two techniques. Besides, they are lightweight and inexpensive, which reinforce their portability (Nakamoto et al., 2018). But their usage still is limited in space and along one plane only. At the moment, an application to accurately measure movements of the back from top to bottom along three axes is still out of reach.

Last, inertial measurement units (IMUs) are electronic devices composed of 2 or 3 sensors (3-axes accelerometer, 3-axes gyroscope and optionally 3-axes magnetometer) that can report the acceleration and orientation of one object using Attitude and Heading Reference System (AHRS) algorithm. They are small and can easily be integrated in a wearable devices (Baek et al., 2010). In addition, they can measure small motions and rotations of the back along the axes with ease (Zhao et al., 2017) and can be combined to tackle a larger array of movements and be more accurate (Chhikara et al., 2008). Last, the physical values obtained from IMUs are very close to the displacements and angles that are usually sought.

Many works emphased the real time features that IMUs can provide in order to design a portable measuring device to gather data on the everyday life of an affected patient to better his treatment and warn him of unsafe positions. Recently, Beange et al. (2019) or Graham et al. (2020) proposed applications of IMU to monition the spine in the context of LBP.

The goal of this study is to provide physician a measuring instrument that could be used to detect and monitor back disorders. As such, the solution must be low-cost, usable in a closed environment without highly-dedicated technical skills. In the following, we will present an IMU-based solution, with detailed validation protocol, and give two first application examples on movement tracking and hip & shoulder dissociation.

## 2. MATERIAL AND METHODS

### 2.1. Device elements

BackMonitor is an in-house system built on the Feather M0 development board (Adafruit Industries[©], New York, USA), an Arduino compatible processor that includes a Bluetooth Low Energy (BLE) module. The sensor (BNO055 Bosch Sensortec - Kusterdingen, Germany) is a 9 DOF (accelerometer, gyroscope, and magnetometer) IMU embedding an AHRS processing. Each sensor has a size of 20mm×27mm×4mm, and weights 3 g; the micro-controller board, can easily to be placed in the trousers pocket (60mm×30mm×25mm, 30g).

### 2.2. Metrology

In the context of low-back pain monitoring, both linear acceleration and Euler angles are useful, respectively for hip & shoulder dissociation detection and for the lumbar lordosis angle follow-up. Those two quantities can be directly gathered from the BNO055 (via an internal fusion algorithm).

BNO055 comes with autocalibration feature. This process is a black box that ought to be verified anyway. Acceleration can be 2-points checked easily by using gravity, but rotation must be studied in more details. Thus, a calibration protractor with IMU holder was 3D printed to set angles with an accuracy of 0.5°.

The angles were measured form each sensor and in each axis between 0° and 165° every 5° for X and Z directions and from 0° to 90° for Y direction according to Euler angle definition. Moreover, the board was initialized while being at 0°.

### 2.3. Assembly

The hardware layout for the simultaneous measurement of data from two sensors is presented in Fig.1, that shows the electronic circuit diagram.

The two BNO055s are connected to the Feather M0 Bluefruit by I2C bus. Each BNO055 has a specific address (respectively 0x29 and 0x28 from left to right) depending on its ADR pin level.

Besides, a 110 mAh battery was followed by a switch connected directly to the processor's power. It is necessary, since the goal is to conceive a portable system. A changeover diode soldered onto the board allows both the USB port and the battery to be connected without any risks as the battery acts as a backup power.

Three switches named A, B and C and a blue LED were added for the user to select the reading mode (hip & shoulder dissociation detection/lumbar lordosis angle measure, serial/BLE connection, raw / AHRS data).

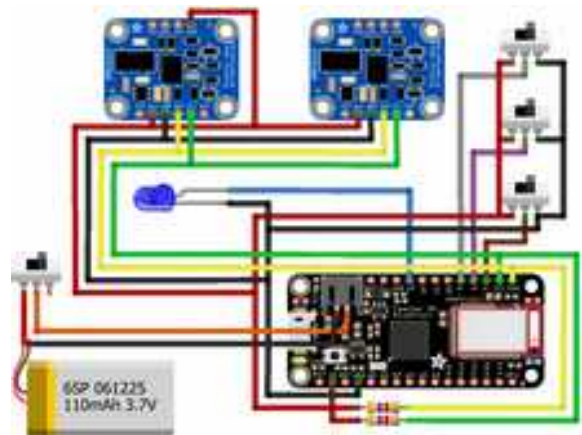

Figure 1. Electrical circuit diagram

### 2.4. Sensors arrangement

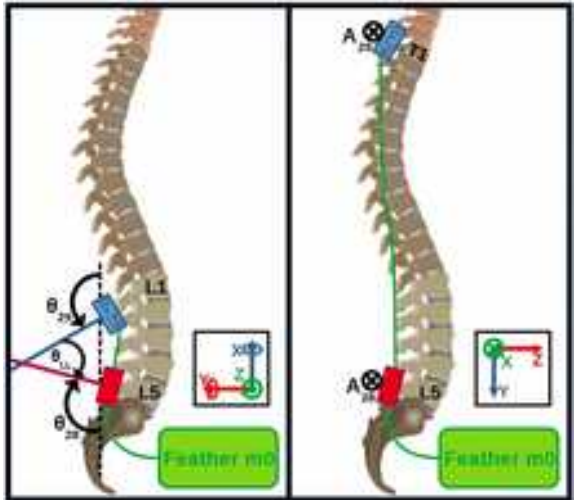

Figure 2. Position of the IMUs for both hip & shoulder dissociation detection (right) and the lumbar lordosis angle measure (left).

The sensor position along the patient's back has been set according to physicians usual practice (Fig.2). Each sensor is stuck on the patient's skin with bio-compatible double-sided adhesive tape.

The lumbar lordosis angle is defined as the angle formed by the tangent to the superior plate of the transitional thoraco-lumbar vertebra the most inclined on the horizontal, usually L1, and the tangent to the inferior plate L5. Thus, one IMU should be put over the L1 vertebra and one over the L5 vertebra. The lumbar lordosis angle $\theta_{LL}$ can be defined as the difference between the Euler angles of the IMUs along the Z axis $\theta_{28}$ and $\theta_{29}$.

Detecting hip & shoulder dissociation is picking up some conflicting accelerations between the upper part of the back and the lower part of the back. As a first suggestion, one IMU is placed over the highest vertebra on the patient's back T1, and one over the lowest L5. The hip & shoulder dissociation will be characterized as the phase difference between the analytic signals obtained from IMU antero-posterior accelerations $A_{28}$ and $A_{29}$ at T1 and L5.

### 2.5. Experimental protocol

A first feasibility test is designed. The system is tested on two volunteers. They are both youth women with the same morphology, one with a medical history of LBP and scoliosis (subject 1) and the other one without any reported back-related issue (subject 2).

Two simple exercises were proposed to the subjects.

First, each subject was asked to sit down on a chair for about 10 seconds, then, to stand up and to stand still for the next 10 seconds. An object was placed 50 cm in front of him and the subject was asked to bend his back in order to pick it up. After that, the patient stood up for 10 seconds once more and then sat down for 10 seconds (Exercise 1).

Second, a time-up-and-go test is done: after 10 seconds on the chair, the patient was asked to walk for about 3 m, turn back and then go back to the chair and stay sit for 10 seconds (Exercise 2).

## 3. RESULTS

### 3.1. Metrology

Table 1 presents the slope and offset between the angle given by the protractor and the measured one. Most of the coefficients of determination $R^2$ are higher that 0.99. The slope is close to 1 – and can be corrected easily. X axis presents the higher sensitivity drift for both sensors.

Table 1. Linearity parameters.

| Sensor name | Axis | offset | Slope | $R^2$ |
|---|---|---|---|---|
| 0x28 | X | -0.5122 | 0.9312 | 0.9995 |
|  | Y | 1.2707 | 0.9844 | 0.9999 |
|  | Z | 1.6358 | 0.9879 | 0.9999 |
| 0x29 | X | -1.9851 | 0.9065 | 0.9973 |
|  | Y | 1.0071 | 0.9692 | 0.9998 |
|  | Z | 6.1829 | 0.9411 | 0.9993 |

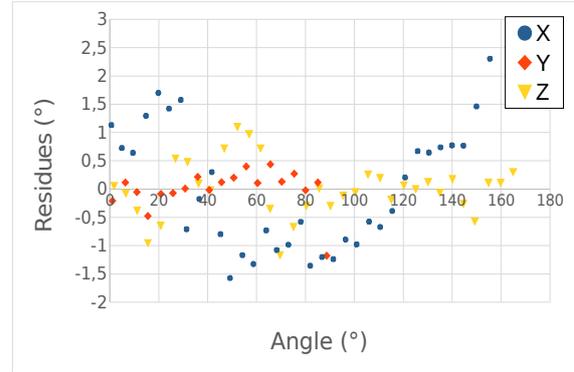

Figure 3. Residues for Euler angles (in degrees) on 0x28

The residue between the best linear fit and the experimental results are the same for 0x28 and 0x29. For Y and Z directions, the residues are essentially random. In the X direction, a $2^{nd}$ order bias appears and can be easily corrected. In these conditions, the standard deviation σ is less than 1° (Fig.3).

The IMUs did not show any measurable drift over a time period of 10 min. Identically, the calibration procedure was performed 10 days after the fist measure, and results were in close agreement. Therefore, it was concluded that reproducibility was good enough and that this point was not an issue.

Last, the BNO055 coming with an auto-calibration feature running after every starting up, it is worth checking the BON055 measurement stability. Measurement at 0° and 75° were performed while varying the starting position. For the Y and Z direction, and for both IMUs, the returned value barely changed with the starting position but it was concluded that the X value was zeroed at the start whatever the real starting angle. Hence, it is required to build a stand for the IMUs so that the initial value for the X axis to be stable for both sensors.

### 3.2. Feasability tests

Fig. 4 presents the sensor 0x28 and 0x29 orientations versus time for the two subjects performing Exercise 1. Sensors are represented by a stick corresponding to their longitudinal axis ($Y^*$). The lumbar lordosis angle is calculated from the angle over the Z axis as follow:

$$\theta_{LL} = -180° - \theta_{28} - \theta_{29} \quad (1)$$

Finally, the lumbar lordosis angle corresponding to each subject was obtained by averaging the values while standing up.

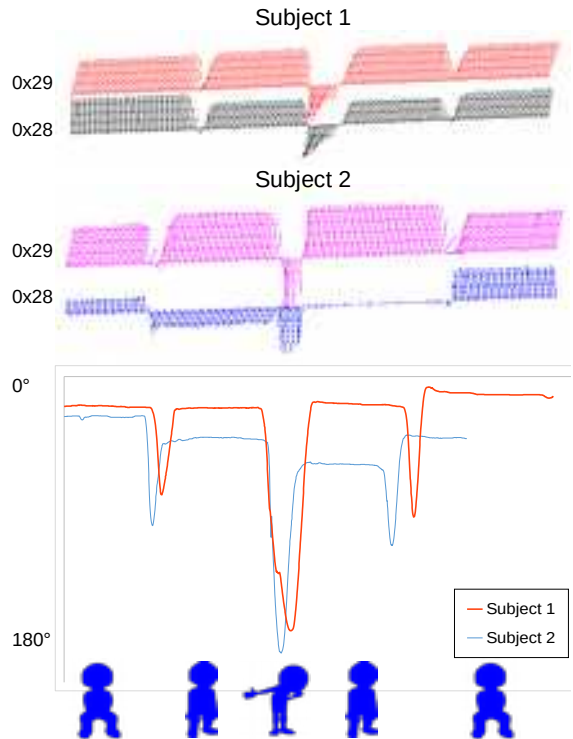

Figure 4. Sensor orientation and lumbar lordosis angle for subject 1 and 2 during Exercise 1.

For the hip & shoulder dissociation detection, charts of linear accelerations along all axis have been produced. The parts where the subject was sat down were remove because they were irrelevant (mostly null) and to better focus on the moving part of the experiment. As the goal is to pick up an offset between the two IMUs, analytical signals on the linear accelerations for the Z axis were calculated with OCTAVE (Eaton, 2019) in order to find raw phase differences (Fig.5).

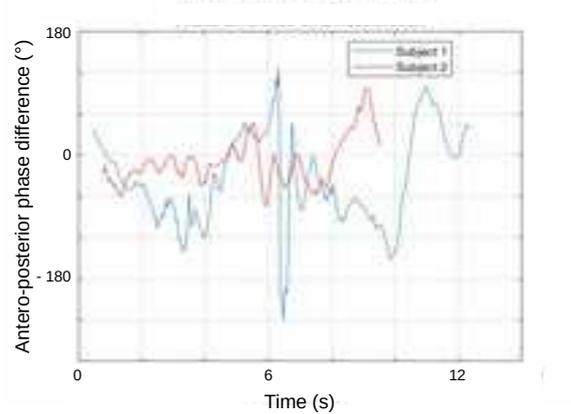

Figure 5. Hip & shoulder phase difference for antero-posterior (Z) acceleration.

## 4. DISCUSSION

The metrological results showed that the sensor arrangement gives reliable results, with a repeatability σ = 1° and no observable drift.

A first test was done, with two women of comparable age and morphology, one with LBP history and the other without. Of course, these results cannot have any statistical meaning; nevertheless, they have a demonstrative interest.

During Exercise 1, the patient afflicted by LBP (subject 1) has a mean lumbar lordosis angle in stand-up phase of −19° and the other −48°. The lumbar lordosis angle being considered as natural in the range −45°±9°, subject 1 is out of the safe interval, unlike subject 2 who is not afflicted by LBP. Fig. 4 shows a difference between the lumbar lordosis angle value in stationary stand-up phase between the 2 subjects. While subject 2's angle varies from −25° to −60°, subject 1's angle stays always close to −20°. Subject 1's movements appear as more restricted in range than subject 2's. More, subject 1's movements are slower than subject 2's.

This is in agreement with previous works stating a decrease both in speed and in range of motion for LBP patients (Errabity et al., 2020).

Exercise 2 focuses on acceleration, and it is possible to extract basic gait analysis information. For example here, subject 1 was about 2s slower than subject 2. But, much detailed observations can be done: while subject 1's steps keep a similar shape and range through time, subject 2's are fluctuating through time. The pain might force subject 1 to limit her walking strategy to few movements while subject 2 can freely adapt her movements to the current stance.

The hip & shoulder dissociation presented as a phase difference on Fig.5 discriminates the two subjects. Indeed, at the turning point (t ≈ 6.5 s), the T1 and L5 vertebrae of subject 1 are nearly in opposition of phase with the lower part of the back lagging behind the upper part. This is not seen in subject 2's case as the phase when turning back (t ≈ 5.5 s) is not much different than when subject 2 is walking. Henceforth, the hip & shoulder dissociation could be detected for the subject with LBP and not for the healthy one using phase analysis.

## 5. CONCLUSION

During this study, we have designed and built a new wearable device capable of detecting features helpful in LBP follow-up while being non-invasive. The metrological validation of BackMonitor arrangement shows good features, with small noise level (σ = 1°) and no observable drift.

Two simple exercises, one combining stand-up, sit and bending movements, the other being a classical time-up-and-go test, were proposed to two young volunteers, one of them with a LBP history. Signal was processed to extract the lordosis angle and hip & shoulder dissociation. Even if no general rules can be extracted from this study, we have shown that IMUs are able to pick up those characteristics and the obtained values are meaningful refereeing to LBP disease.

Hence, we are confident in going to clinical studies to elaborate the link between back related feature and LBP, in particular the hip & shoulder dissociation which is poorly documented.